\documentclass[12pt,preprint]{aastex} 
\newcommand{\be}{\begin{equation}}
\newcommand{\ee}{\end{equation}}
\def\lta{\,\raise 0.3 ex\hbox{$ < $}\kern -0.75 em
 \lower 0.7 ex\hbox{$\sim$}\,}
\def\gta{\,\raise 0.3 ex\hbox{$ > $}\kern -0.75 em
 \lower 0.7 ex\hbox{$\sim$}\,} 
\newcommand{\mplan}{ M_{\rm p}}

\newcommand{\rplan}{ R_{\rm p}}

\newcommand{\mstart}{M_{\scriptstyle I\!I}} 
\newcommand{\mearth}{M_{\oplus}} 
\newcommand{\mjup}{M_J} 
\newcommand{\twig}{{\tilde t}} 
\newcommand{\mdot}{{\dot M}} 
\newcommand{\mdotz}{{\dot M}_0}
\newcommand{\mdotdz}{{\dot M}_{d(0)}}
\newcommand{\mdotd}{{\dot M}_{\rm d}}  
\newcommand{\tcore}{t_{\rm core}} 
\newcommand{\tonset}{t_{\rm onset}} 
\newcommand{\tcorz}{t_{\rm c0}}
\newcommand{\tdisk}{t_{\rm disk}}  
\newcommand{\tend}{t_{\rm end}} 
\newcommand{\fom}{{\cal F}} 

\begin{document} 

\title{A Theoretical Framework for the Mass Distribution of 
Gas Giant Planets forming through the Core Accretion Paradigm} 

\author{Fred C. Adams,$^{1,2}$ Michael R. Meyer,$^2$ and Arthur D. Adams$^2$} 

\affil{$^1$Physics Department, University of Michigan, Ann Arbor, MI 48109} 
\affil{$^2$Astronomy Department, University of Michigan, Ann Arbor, MI 48109} 
\affil{updated: January 2021} 

\begin{abstract}

This paper constructs a theoretical framework for calculating the
distribution of masses for gas giant planets forming via the core
accretion paradigm. Starting with known properties of circumstellar
disks, we present models for the planetary mass distribution over the 
range $0.1\mjup\le\mplan<10\mjup$. If the circumstellar disk lifetime
is solely responsible for the end of planetary mass accretion, the
observed (nearly) exponential distribution of disk lifetime would
imprint an exponential fall-off in the planetary mass function. This
result is in apparent conflict with observations, which suggest that
the mass distribution has a (nearly) power-law form
$dF/d\mplan\sim\mplan^{-p}$, with index $p\approx1.3$, over the
relevant planetary mass range (and for stellar masses
$\sim0.5-2M_\odot$).  The mass accretion rate onto the planet depends
on the fraction of the (circumstellar) disk accretion flow that enters
the Hill sphere, and on the efficiency with which the planet captures
the incoming material. Models for the planetary mass function that
include distributions for these efficiencies, with uninformed priors,
can produce nearly power-law behavior, consistent with current
observations. The disk lifetimes, accretion rates, and other input
parameters depend on the mass of the host star. We show how these
variations lead to different forms for the planetary mass function for
different stellar masses. Compared to stars with masses $M_\ast$ =
$0.5-2M_\odot$, stars with smaller masses are predicted to have a
steeper planetary mass function (fewer large planets).

 
\end{abstract} 

\keywords{planetary systems --- planets and satellites: formation 
--- planets and satellites: dynamical evolution and stability}

\vskip1.0truein 

\section{Introduction} 
\label{sec:intro} 

The mass distribution of fundamental objects represents an important
feature of any astrophysical discipline, including studies of planets,
stars, galaxies, dark matter halos, and galaxy clusters. Moreover, a
complete understanding of the issue requires both the observational
specification of the mass distribution and a predictive theoretical
framework.  The stellar initial mass function (starting with
\citealt{salpeter}), for example, affects a large number of
astronomical issues, including galactic chemical evolution, supernova
rates, and feedback during star formation. With thousands of
exoplanets now discovered (e.g., see \citealt{exoorg,gaudi2020}), we can 
start to address the planetary mass function (PMF), which will ultimately
inform studies of planetary system architectures, planetary habitability,
and many other issues. In addition, the PMF will provide a consistency
check on theories of planet formation, once they are sufficiently
developed. The observational determination of this mass distribution
is now coming into focus, although the database remains incomplete,
and many uncertainties persist.  On the other hand, our theoretical
understanding of the mass function remains in its infancy. The goal of
this paper is to construct simple but physically motivated models for
the PMF for giant planets, i.e., companions in the mass interval
$\mplan=0.1-10\mjup$ (where $\mjup$ is the mass of Jupiter).

For the range of planetary masses of interest, most objects are
thought to form via the core accretion paradigm \citep{pollack}, which
is organized into three phases (see, e.g., \citealt{benz2014} for a
review). In the context of this mechanism, the accumulation of small
rocky bodies (phase I) leads to the production of planetary cores with
masses of order $10\mearth$, which takes place over a time scale
$\tcore$. Note that core formation could take place either through
two-body accumulation of icy planetesimals \citep{hansen2013} or
through rapid accretion of small pebbles assisted by gas drag
\citep{ormelklahr,lambrechts12}. In any case, the cores accrete gas
from the surrounding disk after reaching their critical mass
threshold. During the subsequent phase II, growth is relatively slow
and is controlled by cooling processes (e.g., opacity effects). After
the total mass reaches roughly twice the core mass (mass scale
$\mstart$) at time $\tonset$, gas accretion occurs rapidly and the
planets accumulate most of their mass. This phase III lasts until the
planets reach their final masses in the range
$\mplan\approx0.1-10\mjup$, and then terminate their growth at time
$\tend$. The final mass of a given planet is thus given by the
integral expression 
\be 
\mplan = \mstart + \int_{\tonset}^{\tend} \mdot_{\rm p}(t)\, dt \,\,
= \mstart + \langle\mdot_{\rm p}\rangle (\tend-\tonset) \,\,,
\label{mbasic} 
\ee
where $\mdot_{\rm p}(t)$ is the accretion rate onto the planet as a
function of time, and where the final equality follows from the mean
value theorem. Within the context of the core accretion paradigm, the
mass accretion rate onto the planet is some fraction of the total mass 
accretion rate $\mdotd$ through the disk, so that 
$\langle\mdot_{\rm p}\rangle$ = $\langle{f}\mdotd\rangle$ $\equiv$
$\epsilon \langle\mdotd\rangle$, where the final equality defines a
time-averaged efficiency factor $\epsilon$. Since circumstellar disks
have lifetimes $\sim1-10$ Myr, it is natural to associate the end of
planetary accretion ($\tend$) with the disappearance of the disk. 
One objective of this paper is to test the plausibility of this 
hypothesis. 

As written, equation (\ref{mbasic}) is {\it exact}, and depends only
on the four variables ($\mstart$, $\langle\mdot_{\rm p}\rangle$,
$\tonset$, $\tend$). In other words, for a particular planet-forming
event, if we know the mass $\mstart$ at the onset of gas accretion, the
starting and ending times $(\tonset,\tend)$, and the mean accretion
rate $\langle\mdot_{\rm p}\rangle$, the final mass of the planet is
completely determined. If, in addition, we know the distributions of
the variables over the collection of planet forming disks of interest,
then the distribution of planetary masses is also determined.
Unfortunately, the mass accretion history of the planet, along with
the time scales $\tonset$ and $\tend$, depend on a large number of
physical processes that remain under study, including disk viscosity
which drives accretion, the fraction of material moving through the 
disk captured by the growing planet, disk and planetary magnetic
fields, planetary migration, the role of circumplanetary disks, and
many others. In a complete theory, where all of the relevant
subprocesses are known and predictable, one could calculate $\tonset$,
$\tend$, and $\mdot_{\rm p}(t)$, and their distributions, from first
principles.  In the absence of such a complete understanding, the more
modest goal of this paper is to construct observationally motivated
distributions of the relevant input variables, as encapsulated via
equation (\ref{mbasic}), and use the results to predict the PMF. We
can then determine the properties of these input distributions
required to produce planetary mass functions that are consistent with
observations. In addition, the framework developed herein can be
readily generalized in future work.

Although no rigorous, {\it a priori} theory of the planetary mass
function has been established, a substantial amount of previous work
has been carried out. One approach is to use population synthesis
models, wherein planet formation is simulated using a large number of
physical processes that are treated in an approximate manner (e.g.,
see \citealt{idalin2008,benz2014}). With the proper choice of input
parameters, these models produce mass distributions for planets that
are roughly consistent with the observed PMF
\citep{thommes2008,mordasini2009}, although they tend to underproduce
sub-saturn mass planets \citep{suzuki2018}. The approach of this paper
is complementary to population synthesis models. Instead of including
as many physical processes as possible, our goal is to find the
minimal level of complexity necessary to reproduce the observed PMF.
This present approach is semi-empirical in that it relies on
observational input to help specify the distributions of input
parameters (see \citealt{adamsfat} for an analogous approach to the
stellar initial mass function). In addition, this minimalistic
approach allows for PMF expressions to be obtained semi-analytically.
Other related studies include a statistical model based on orbital
spacing \citep{malhotra2015} and the role of dynamical instabilities
in shaping the planetary mass function \citep{carrera2018}.

The scope of this paper is limited to gaseous giant planets forming
through the process of core accretion, in particular those objects for
which the gas accretion phase determines the final planetary mass.  
As a result, this approach is limited to planets forming in the mass
range $0.1\mjup\le\mplan\le10\mjup$. The current observational sample,
especially from the {\it Kepler} mission \citep{borucki,batalha},
indicates that the total planet population includes a large number of
smaller bodies, with a preference for superearths with masses
$\mplan\sim10\mearth$ (e.g., \citealt{zhu2018}). These objects are
roughly analogous to the cores of the giant planets, but they are
thought to contain relatively little gas, and their mass distribution
is not addressed here. These lower mass planets could represent a
second population (\citealt{pascucci}; see also
\citealt{schlaufman2015}), which should be addressed in future
studies. In addition, this present analysis is confined to those
star/disk systems where giant planets are able to form. The
probability that a given system will produce giant planets is a
separate but important issue \citep{howard2010,newmeyer}.

This paper is organized as follows. Section \ref{sec:observe}
discusses the observational constraints utilized in our models,
including a brief overview of the observed planetary mass function,
distributions of disk lifetimes, constraints on core formation time
scales, and the dependence of these quantities on the mass of the host
star. Section \ref{sec:models} presents a framework to construct the
PMF. A model using the observed distribution of disk lifetimes, the
expected dependence of the mass accretion rate on the Hill radius, and
an additional random variable can reproduce the nearly power-law mass
function that is observed. Since disk properties, and hence planet
properties, depend on the mass of the central star, Section
\ref{sec:varymass} explores the effects of varying the stellar mass.
The paper concludes in Section \ref{sec:conclude} with a summary of
our results and discussion of their implications. For completeness,
the paper includes an appendix that considers an alternative model for
the PMF.

\section{Empirical Considerations} 
\label{sec:observe} 

This section outlines the observational input used to inform and test
our models of the planetary mass function. We start with current
constraints on the observed PMF. Unfortunately, the observed mass
distribution function for gas giant planets is not fully specified
with current data. Studies carried out to date for FGK stars (starting
with \citealt{cumming2008}; see also \citealt{mayor2011}) suggest that
the mass distribution has an approximately power-law form over the
mass range of interest $0.1\mjup\le\mplan\le10\mjup$, i.e., 
\be
{dF \over d\mplan} = {A \over \mplan^p} \qquad {\rm where} \qquad 
p \approx 1.3 \pm 0.2 \,,
\label{mpower} 
\ee 
where $A$ is a normalization constant (and more data are needed at
the ends of the mass range). Although significant uncertainties
remain, this paper uses the power-law mass distribution of equation
(\ref{mpower}) as its basis for comparison to observations. Note that
this form for the mass function is only applicable over the specified
range of planetary masses. The distribution of planet radii displays a
break near $\rplan=4R_\oplus$ \citep{schlaufman2015}, corresponding to
planet mass $\mplan$ = $10-15\mearth$ \citep{wolfgang}, which falls
just below the masses of interest here. Extending the range to smaller
masses, down to $\mplan\sim1\mearth$, the planetary radius
distribution has a bimodal form (e.g., \citealt{fulton2017}). Another
complication is that planets of a given mass can display a
distribution of radii \citep{otegi2020}.  Current results from
microlensing surveys \citep{suzuki2016,shvartzvald} also find
power-law distributions for the ratios of companion masses
$q=M_2/M_\ast$ over the range $10^{-4}<q<10^{-2}$, with slopes
$p\approx0.9-1.5$, roughly consistent with equation (\ref{mpower}).
Keep in mind that these surveys primarily sample host stars of lower
mass (which are the most common stars).  For completeness, we note
that the observed mass distribution for wide-orbit planets with masses
$\mplan>3\mjup$ follows a similar power-law
\citep{schlaufman2018,wagner2019}, although the Gemini Planet Imager
Exoplanet Survey hints at a steeper slope $p\approx2.3\pm0.8$
\citep{nielsen2019}.

One should keep in mind that the observed PMF is derived from planets
detected around older main sequence stars, and this distribution could
have evolved from earlier epochs immediately after planet formation.
For example, planet-planet interactions can eject planets from their
solar systems, and migration through the disk can lead to accretion of
planets by their host stars. Since these processes are unlikely to be
independent of planetary mass, their action will sculpt the PMF over
time.  These processing effects, and perhaps others, should thus be
included in the determination of the observed PMF. In other words, we
ultimately need to make the distinction between the initial planetary
mass function and the observed planetary mass function (an analogous
complication is well known in considerations of the stellar mass
distribution).

The disk lifetime sets an upper limit on the time available for gas
accretion onto planets. The disks associated with young stars
generally have lifetimes in the range 1 -- 10 Myr
\citep{meyer2007,williams,hartmann2016,martinez}. More specifically,
the distribution of disk lifetimes $dF/dt$ is often taken to have the
general form 
\be 
{dF \over dt} = {1 \over \tau} \exp(-t/\tau) \,,
\label{tdist} 
\ee 
which is consistent with observations of disk fractions in star
forming regions of varying ages
\citep{haisch2001,jesus,mamajek,fedele2010}. Note that different
signatures are used to indicate the presence of disks, including signs
of active gas accretion onto the stars, near-infrared excess emission
at $1-5\mu$m (tracing the inner disk), infrared emission at
$5-100\mu{m}$ (tracing disk scales 0.3 -- 30 AU), and millimeter-wave
observations of dust and gas (tracing the outer disks at $10-100$ AU). 
As written, the distribution of disk lifetimes is normalized such that 
\be 
\int_0^\infty {dF\over dt} dt = 1 \,.  
\label{tnorm} 
\ee
Observations indicate that the time scale $\tau\approx4-5$ Myr. Note
that $\tau$ represents the time required for the population of disks
to decrease by a factor of e $\approx2.72$. The corresponding
`half-life', the time required for the population to decrease by 
a factor of 2, is only about $\tau_{1/2}=\tau\ln2\sim3$ Myr. 

Observations also indicate how disk properties vary with the mass of
the host star. Disk lifetimes are observed to decrease with increasing
stellar mass \citep{hillenbrand,carpenter,yasui,ribas2015,yao2018}. 
This observational finding can be incorporated using a scaling law of
the form 
\be
\tau = {\tau_1 \over m} \,,
\label{tauscale} 
\ee
where $\tau_1\sim5$ Myr, and where we have defined a 
dimensionless stellar mass variable 
\be
m \equiv {M_\ast \over 1\,\,M_\odot} \,.
\label{mdef} 
\ee 
Mass accretion rates $\mdotd$ for the disks surrounding T Tauri
stars depend on both time and the mass of the central star. 
These dependences can be modeled with a function of the form 
\be
\mdotd =\mdotdz {m^2 \over (1 + t/t_0)^{3/2}} \,, 
\label{mdotdisk} 
\ee
where the time scale $t_0\sim1$ Myr and where the time dependence
follows from $\alpha$-disk models (e.g., \citealt{leebook}).  The
observed $m^2$ dependence (in the numerator) follows from
observational results (see the review of \citealt{hartmann2016}).
Although no simple explanation is currently accepted for this law, 
it follows if disk lifetimes decrease with stellar mass (equation
[\ref{tauscale}]) and the starting disk mass $M_{\rm d}\propto
M_\ast$ increases with stellar mass (see also 
\citealt{hartmann2006,dullemond2006}). This latter scaling law 
has been observed in a number of surveys (e.g.,
\citealt{natta2007,andrews2013,ansdell2016}), with the slope somewhat
steeper than linear in some studies (e.g.,
\citealt{pascucci2016,barenfeld2016}). In general, the observations
find a well-defined correlation between submillimeter disk luminosity
and stellar mass, whereas the conversion from the observed quantities
to the inferred relationship $M_{\rm d}\propto M_\ast$ requires disk
modeling. In addition, the observations correspond to dust emission,
so that the dust-to-gas ratio must be assumed in order to obtain
estimates for the total disk mass. For completeness, note that a
weaker correlation is found for actively evaporating disks in the
Orion Nebula Cluster \citep{eisner2018}.

Finally, the formation time $\tcore$ for the cores of giant planets,
and hence the time $\tonset$ before the start of rapid gas accretion,
is also expected to depend on the mass of the central star. Here we
adopt a scaling law of the form 
\be
\tonset = \tcorz\,m^{1/2} \,, 
\label{tcorescale} 
\ee
where we expect that $\tcorz\sim1$ Myr \citep{pollack,benz2014}.
This scaling law follows from theoretical considerations of solid
accumulation \citep{safronov} combined with empirical scaling laws. 
We start by assuming that planetary cores preferentially form just
outside the water iceline, where the surface density of solids is
expected to increase.  Using the mass-luminosity relation for
pre-main-sequence stars,\footnote{The index depends on the specific
evolution time and mass range considered. This result follows from
considerations of Hayashi's forbidden region (e.g., \citealt{hansen}),
and can also be determined from stellar evolution simulations (e.g.,
\citealt{paxton}).} $L_\ast\sim M_\ast^\mu$, where $\mu\approx5/3-2$,
and a fixed sublimation temperature $T_{\rm ice}$, we can find the
semi-major axis of the iceline from energy balance. 
Since $T_{\rm ice}^4\propto L_\ast/a^2$ $\propto M_\ast^2/a^2$ = 
{\sl constant}, we find that $a_{\rm ice}\propto M_\ast$. The larger
radius of the formation site --- for larger stars --- leads to 
slower accumulation of the cores.  The Safronov accumulation time
(specifically, the doubling time in the absence of gravitational
focusing) depends on the surface density $\Sigma$ and orbital
frequency $\Omega$ (e.g., \citealt{safronov77}) according to 
$t_{\rm saf}\propto\Sigma^{-1}\Omega^{-1}$. Using a surface density
$\Sigma\propto M_\ast/a^{1/2}$ (e.g., \citealt{andrews2009}), along
with a Keplerian rotation curve, we find the result given in equation
(\ref{tcorescale}). Note that steeper surface density profiles lead to
more sensitive dependence of the core formation time on stellar mass.

The particular scaling law (\ref{tcorescale}) was derived for the case
of classic planetesimal accretion. For the competing picture wherein
pebble accretion accounts for the formation of giant planet cores
(e.g., \citealt{ormelklahr,lambrechts12,lin2018,rosenthal}), the
scaling exponent can be different. For example, \cite{lambrechts14}
find that the pebble accretion rate depends on stellar mass according
to ${\dot M}\sim M_\ast^{-11/36}$, which coresponds to the alternate
scaling law $\tonset\sim m^{11/36}$ $\sim m^{1/3}$ (compare with
equation [\ref{tcorescale}]). The goal of this present work is to
explore the effects due to the core formation time increasing with
stellar mass, but the exact scaling is not as important.  For
completeness, we note that pebble accretion can occur over a wider
range of radial locations in the disk, compared with the classical
picture.\footnote{We also note that the migration of solid material
could affect this scaling law, if the migration rates depend on the 
stellar mass.} Nonetheless, studies suggest that pebble accretion is
more efficient beyond the ice line, as implicitly assumed here, due to
the greater mass in pebbles and higher probability of sticking (e.g.,
\citealt{bitsch,chambers2016}).

\section{Planetary Mass Function for Exponential Distribution 
of Disk Lifetimes} 
\label{sec:models} 

This section constructs a working model for the planetary mass
function based on the empirical considerations of the previous
section. The models constructed here are independent of the mass of
the host star. The effects of varying stellar mass on the resulting
PMFs are addressed in Section \ref{sec:varymass}.

\subsection{Distribution of Gas Accretion Time Scales} 
\label{sec:tdecay} 

Observations provide estimates for the distribution of disk lifetimes.
For purposes of determining planetary masses, however, we need the
distribution of time remaining after the onset of rapid gas accretion.
The starting time $\tonset$ is determined by the core formation time
scale and the slow cooling phase. It takes into account the fact that
rapid gas accretion onto the planet occurs only after the body has
built up a critical mass, which includes both the rocky core and
enough additional gas to become self-gravitating. For a given stellar
$M_\ast$, the time to build the core $\tonset$ depends on the surface
density of solids and cooling time depends on the gas opacity in the
planetary atmosphere (e.g., see 
\citealt{pollack,benz2014,helled2014,piso2015}, and references therein).

For the case of interest here, where disk lifetimes are exponentially
distributed, the remaining accretion time also follows an exponential
law. Consider a given value $\tonset$ of the starting time and define
a new effective time variable $\twig$ according to 
\be
\twig = t - \tonset \,,
\label{twigdef} 
\ee
which resets the zero point of time for gas accretion. The time $t$ is
the disk lifetime, which represents the time when accretion ends, and
which follows the exponential distribution of equation (\ref{tdist}).
For any value of $\tonset$, the probability distribution of the
corrected time variable $\twig$ is determined by 
\be
{dF\over d\twig} = {dF\over dt} {dt\over d\twig} = 
{1\over\tau} \exp[-(\twig+\tonset)/\tau] = 
{\exp[-\tonset/\tau]\over\tau} \exp[-\twig/\tau] \,,
\label{twigzero} 
\ee
where $dF/dt$ is the distribution for the original time variable $t$.
The probability distributions for both $t$ and $\twig$ thus have the
same exponential form.  As written, however, this expression for the
probability distribution (for $\twig$) is not normalized. Its integral
over all $\twig>0$ is less than unity because the leading coefficient
takes into account the fact that those disks with lifetimes
$\tdisk<\tonset$ will not produce gas giant planets. For purposes of
calculating probability distributions for the time remaining after the
onset of rapid gas accretion, however, the normalized form of the 
distribution must be used, i.e.,  
\be
{dF \over d\twig} = {1\over\tau} \exp[-\twig/\tau] \,.
\label{twigdist} 
\ee
After disks with short lifetimes are removed from the sample, the
remaining disk population will `continue to decay' with an exponential
law that has the same `half-life' as before (analogous to the case of
radioactive decay). 

If all of the disks produced planetary cores on the same time scale,
then the factor $\exp[-\tonset/\tau]$ appearing in equation
(\ref{twigzero}) would determine the fraction of star/disk systems
that could produce Jovian planets. In the absence of additional
mechanisms that suppress planet formation, the expectation value
$\langle\exp[-\tonset/\tau]\rangle$ is proportional to the planetary
occurrence fraction. In practice, however, the timescale $\tonset$
depends on stellar mass, so the corresponding planetary occurrence
fraction is also mass dependent. In addition, the mass accretion rate
depends on stellar mass, so the resulting masses of the planets that
form also depend on $M_\ast$ (see Section \ref{sec:varymass}). 
Note that the observed occurrence rate for Jovian planets
($\mplan=1-10\mjup$), while highly uncertain, is estimated to fall 
in the range 10 -- 30 percent for FGK stars (integrated over all 
separations; see \citealt{winnfab,newmeyer}, and references therein).
Consistency implies that the typical delay time for the onset of rapid
gas accretion is bounded by $\tonset\lta\tau\sim5$ Myr. This estimate
is compatible with estimates from core accretion models
\citep{pollack,benz2014}, although giant planet formation is likely to
face additional bottlenecks, and this issue should be explored
further.

\subsection{Mass Accretion Rate onto the Planet} 
\label{sec:mdot} 

The mass accretion rate onto a forming planet represents a fundamental
variable of the problem. We can conceptually divide the process into
three components. [a] On the largest scale of the circumstellar disk,
an accretion flow moves mass inward at a rate $\mdotd$.  Some fraction
of this mass flow will enter the sphere of influence of the planet,
where this fraction can vary from system to system. [b] Since the
sphere of influence of the planet grows with planetary mass, the mass
accretion rate onto the planet is expected to increase as the planet
becomes more massive.\footnote{Strictly speaking, this statement holds  
provided that the disk properties do not change rapidly enough to
compromise the mass supply to the planet. Since mass accretion rates
through the disk can be episodic, for example, the mass accretion rate
onto the planet can be non-monotonic. }  [c] Only some fraction of the
material that enters the Hill sphere will actually be accreted onto
the planet. This treatment includes these factors as decribed below:

For the latter stages of gas accretion, the rate of accretion onto the
planet depends on the Hill radius $R_H$ raised to some power, where
\be
R_H = a \left({\mplan \over 3 M_\ast}\right)^{1/3} \,, 
\label{rhill} 
\ee where $a$ is the semimajor axis of the forming planet and $\mplan$
is its instantaneous mass.  The fiducial cross section scales as
$R_H^2$ \citep{zhu2011,durmann2017}.  However, numerical simulations
indicate that the mass accretion rate onto forming planets scales
according to steeper power-law $\mdot \sim R_H^4$. The additional
factor of $R_H^2$ arises from the shock of the inward flowing gas at
the boundary defined by the Hill radius \citep{tanwat2002,tantan2016}.
The shock raises the density by a factor of $(v/v_s)^2$, where the
incoming speed $v$ is given by $v\sim\Omega R_H$ and $v_s$ is the
sound speed (see also \citealt{lee2019}). As a result, the mass
accretion rate must scale with the mass of the forming planet \be
\mdot \sim R_H^4 \sim M_p^{4/3} \, .  \ee If additional processes are
operational, then accretion onto the planet can be suppressed
further. Possible mechanisms that come into play include suppression
by gap opening (which limits the density of the incoming flow;
\citealt{malik,ginzburg2019}), strong magnetic fields on the planetary
surface \citep{batygin2018,cridland2018}, suppression of direct infall
due to the initial angular momentum of the incoming material
\citep{machida}, and processes taking place within circumplanetary
disks \citep{szulagyi,fung2019}. The latter three processes are
analogous to well-studied mechanisms involved in the star formation
process. Strong stellar magnetic fields truncate the disk and shut
down direct accretion, which occurs along field lines at an attenuated
rate. The initial angular momentum of the pre-collapse gas causes most
of the incoming material to fall initially onto a circumstellar disk, 
rather than directly onto the star. Although these processes are not
well-studied in the context of planet formation, they will act to
suppress mass accretion and are likely to vary from source to source,
thereby introducing a random variable into the problem.

Putting all of the components together, we can write the mass 
accretion rate onto the planet in the form 
\be
\mdot = \xi \mdotz \left( {\mplan \over \mjup} \right)^{4/3} \,.
\label{mdothill} 
\ee
In this form, the benchmark value of the mass accretion rate is that
expected when the planet has a Jovian mass.  In order of magnitude,
for typical disk properties and accretion efficiencies, we expect
$\mdotz\sim1\mjup$ Myr$^{-1}$, but a range of values are possible
\citep{helled2014}. The leading coefficient $\xi$ is a random variable
on [0,1] that takes into account both the efficiency for material to
enter the Hill sphere and the efficiency with which the planet
accretes the incoming material. In principle, one could keep these two
efficiencies as separate variables. In this treatment, however, we
consider only a single random variable $\xi$. Moreover, given that the
mass accretion process has a well-defined scale set by the mass
accretion rate through the disk, the natural starting point is to
consider the random variable $\xi$ to have a uniform
distribution.\footnote{In contrast, for problems without a  
well-defined fundamental scale, the natural choice for the probability
density function is log-uniform, or $dP/dx\propto1/x$
\citep{jefferys}, which is often called the {\it Jefferys prior}.}

The mass accretion rate onto the planet from equation (\ref{mdothill})
is a rapidly increasing function of planetary mass. Once the planet
grows sufficiently large, the rate of accretion onto the planet can
become comparable to the rate of mass accretion through the disk. At
this point in time, $\mdot$ must saturate, and would no longer follow
equation (\ref{mdothill}).  For typical parameters ($\xi=1/2$ and
$\mdotz=1\mjup$ Myr$^{-1}$; e.g., \citealt{helled2014}), and a disk 
accretion rate $\mdotd=10^{-8}M_\odot$ yr$^{-1}$ (e.g., 
\citealt{gullbring,hartmann2016}), this crossover point corresponds to
planetary mass $\mplan\approx10\mjup$.  Larger disk accretion rates
require larger planetary masses to reach saturation (which would depend
on stellar mass). Since these mass scales are outside the range of
interest, and since including this effect requires specification of
additional parameters, we do not model the saturation of $\mdot$ in
this present paper. Nonetheless, if the saturation of the mass
accretion rate takes a specified form, it is straightforward to
incorporate this complexity. Finally, we note that including this
effect leads to lower mass accretion rates for sufficiently large
planets, thereby leading to somewhat lower masses.

\subsection{Probability Distribution Function} 
\label{sec:pdf} 

Starting with the form for the mass accretion rate from equation 
(\ref{mdothill}), we can integrate over time to find an expression 
for the planet mass
\be
\mplan = \mstart \left[ 1 - \xi 
{\mstart^{1/3} \mdotz \twig \over 3\mjup^{4/3}} \right]^{-3} \,.
\label{semf} 
\ee
The scale $\mstart$ is the mass of the planet at the end Phase II
(start of Phase III), when gas accretion takes place rapidly. For the
sake of definiteness, we take $\mstart=20\mearth$ (about twice the
core mass).  The expression (\ref{semf}) thus contains two random
variables $(\xi,\twig)$. The cummulative probability that one will
draw the two variables in order to obtain a planet mass less than
$\mplan$ is given by the integral 
\be
P (m<\mplan) = \int_0^{t_\ast} 
{d\twig\over\tau} \exp[-\twig/\tau] + 
\int_{t_\ast}^\infty {d\twig\over\tau} \exp[-\twig/\tau] 
{t_\ast \over \twig} \,,
\ee
where we have defined 
\be
t_\ast \equiv {3\mjup^{4/3} \over \mstart^{1/3} \mdotz} 
\left[ 1 - \left({\mstart\over\mplan}\right)^{1/3}\right] \,. 
\label{tstar} 
\ee
The time scale $t_\ast(\mplan)$ is the time required to build a planet
of mass $\mplan$ at the maximum rate (with $\xi=1$).  The differential
probability (probability density function) is thus given by 
\be
{dP \over d\mplan} = {1 \over \tau \mdotz} 
\left({\mjup\over\mplan}\right)^{4/3} 
\int_{t_\ast}^\infty {dt \over t} \exp[-t/\tau] 
= {1 \over \tau \mdotz} 
\left({\mjup\over\mplan}\right)^{4/3} 
E_1(t_\ast/\tau) \,,
\label{eonedist} 
\ee
where $E_1(z)$ is the exponential integral \citep{abrasteg} and where
$t_\ast$ is given by equation (\ref{tstar}). Note that only the
product $\mdotz\tau$ of the fiducial mass accretion rate and time
scale $\tau$ of the disk lifetime distribution appear in the final
expression. To evaluate the above expressions, it is thus useful to
define a composite parameter -- which is a mass scale --- according to 
\be
M_0 \equiv \left( {\mstart \over \mjup} \right)^{1/3} 
\mdotz \tau \,.
\label{mzero} 
\ee
For typical parameter values ($\mstart=20\mearth$, $\mdotz\sim2\mjup$
Myr$^{-1}$, and $\tau=5$ Myr), we find the scale $M_0\approx4\mjup$. 
Given that the parameters are uncertain, we can find the distributions
of planet masses with varying values of $M_0$. Note that the values of
the starting mass $\mstart$ and the time scale $\tau$ of the disk
lifetime distribution are relatively well known. As a result,
specification of the mass scale $M_0$ is largely determined by the
baseline value of the mass accretion rate $\mdotz$.

\begin{figure}
\includegraphics[scale=0.70]{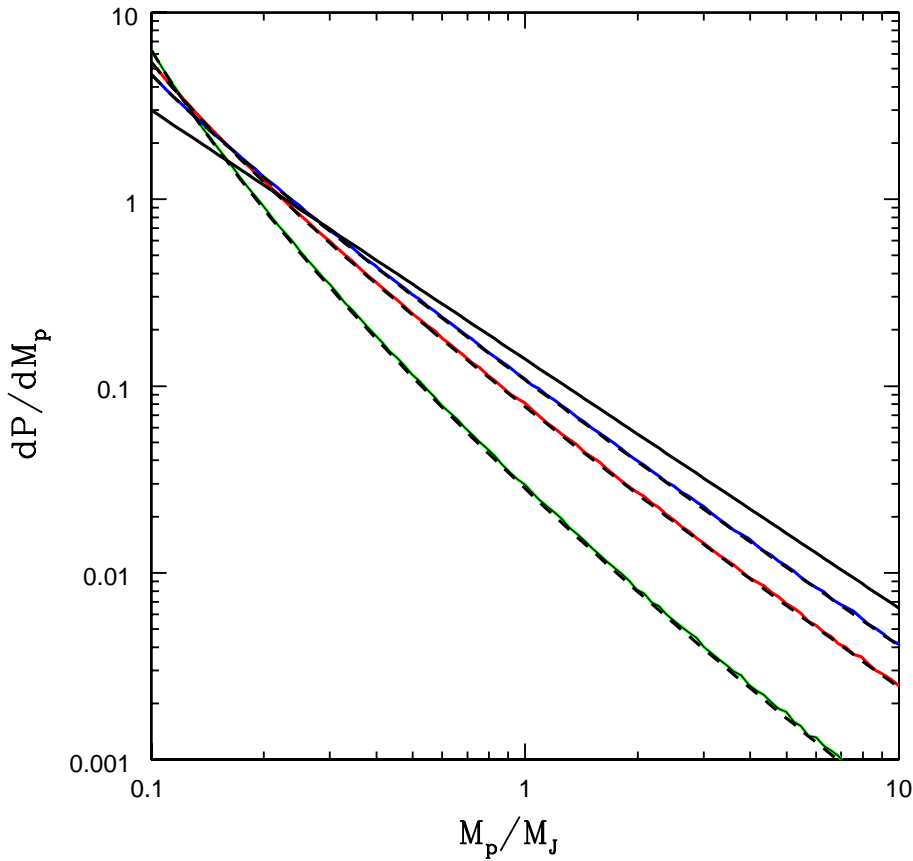}
\vskip-1.50truein
\caption{Planetary Mass Function. The curves show the planetary
mass function predicted from an exponential distribution of disk
lifetimes and a uniform-random distribution of mass accretion rates.
The distributions are characterized by the mass scale $M_0$, which is
determined by the the overall mass accretion rate $\mdotz$ and the
time scale $\tau$ of the disk lifetime distribution.  Results are
shown for $M_0=1\mjup$ (green) $\sqrt{10}\mjup$ (red), and $10\mjup$
(blue). For each case, the colored curves are determined by sampling
from the distributions, whereas the underlying dashed black curves 
show the analytic result from equation (\ref{eonedist}). For
comparison, the solid black curve shows a power-law mass distribution
($dP/dM\sim M^{-1.3}$). }
\label{fig:pmfe1} 
\end{figure}

Given the above results, we can construct planetary mass functions in
two ways: First, we can sample the variables in equation (\ref{semf})
using their specified distributions, calculate the masses, and
construct the histogram of results. Second, we can use equation
(\ref{eonedist}). Both results are shown in Figure \ref{fig:pmfe1} for
different choices of the fiducial mass scale in the range $M_0$ = 1 --
10 $\mjup$. The sampling results are shown as the colored curves and
the analytic results are shown dashed black curves (note that the
results are essentially the same).  For comparison, the solid black
curve shows the expected power-law mass function of the form
$dF/dM\propto M^{-1.3}$.  With low accretion rates (mass scale)
$M_0=1\mjup$ (green curve), the model produces a PMF that is steeper
than the target distribution, with a deficit of high mass planets. 
For larger values of $M_0$, the theoretical PMF approaches a power-law
form over most of the relevant mass range and is in reasonable
agreement with the target distribution. Nonetheless, the theoretical
distributions remain slightly steeper.

As mentioned above, the mass accretion rate from equation
(\ref{mdothill}) is an increasing function of time. If the planet mass
become sufficiently large, the predicted mass accretion rate becomes
larger than the rate supplied by the circumstellar disk, so that the
planetary mass accretion rate must saturate and approach a constant
(maximum) value. The PMF here is constructed using the increasing form
of the mass accretion rate. This assumption leads to somewhat larger
planets for the tail of the distribution. For completeness, we also
consider the opposite limit in Appendix \ref{sec:appendix}, where the
mass accretion rate saturates and thus becomes constant while most of
the planetary mass is acquired. This Appendix also illustrates how the
PMF framework constructed in the paper can be modified or generalized.

\section{Planetary Mass Functions for Varying Stellar Mass} 
\label{sec:varymass} 

The treatment thus far does not include the mass of the host star in
the analysis. This section uses the observed properties of
circumstellar disks, which depend on the mass of the star, to inform
the models of the planetary mass function constructed in the previous
section.  Within this framework, the PMF is specified up to the
characteristic mass scale $M_0$ defined in equation (\ref{mzero}).
Here we use the empirical scaling laws outlined in Section
\ref{sec:observe} to define how the mass scale $M_0$ varies as a
function of the stellar mass, and use the result to specify the PMF
for varying $M_\ast$.

Consider the accretion of gas onto a forming giant planet where the
rate is a constant fraction $\epsilon$ of the disk accretion rate.
Including the time dependence of the disk mass accretion rate from
equation (\ref{mdotdisk}), the total mass accreted --- and hence the
mass of the planet --- must be given by the integral 
\be
\mplan = \int_{\tonset}^{\tend} \epsilon \mdotdz 
{m^2 dt \over (1 + t/t_0)^{3/2}} = 2 \epsilon \mdotdz t_0 
m^2 \left[ {1 \over (1 + \tonset/t_0)^{1/2}} - 
{1 \over (1 + \tend/t_0)^{1/2}} \right] \,,
\ee
where $m$ is the dimensionless stellar mass from equation (\ref{mdef}).
In order to define the characteristic mass scale $M_0$ as a function
of $m$, we use the scaling laws from Section \ref{sec:observe}. The
core formation time varies with stellar mass according to equation
(\ref{tcorescale}). The typical time for accretion to end is given by
the disk lifetime parameter $\tau$, which varies with $m$ according to
equation (\ref{tauscale}). We can then define the mass scale $M_0$
using the ansatz 
\be
M_0 = \left[ \epsilon \mdotdz t_0 \right] \fom (m) 
\qquad {\rm where} \qquad 
\fom (m) \equiv 2m^2 \left[ {1 \over (1 + \sqrt{m})^{1/2}} - 
{1 \over (1 + 5/m)^{1/2}} \right] \,. 
\label{massint} 
\ee 
Notice that the function $\fom(m)\to0$ for stellar mass $m$ =
$(\tau_1/\tcorz)^{2/3}\sim3$. Our interpretation of this finding is
that relatively few planets should form in disks associated with
larger stellar masses (see the discussion below).

\begin{figure}
\includegraphics[scale=0.70]{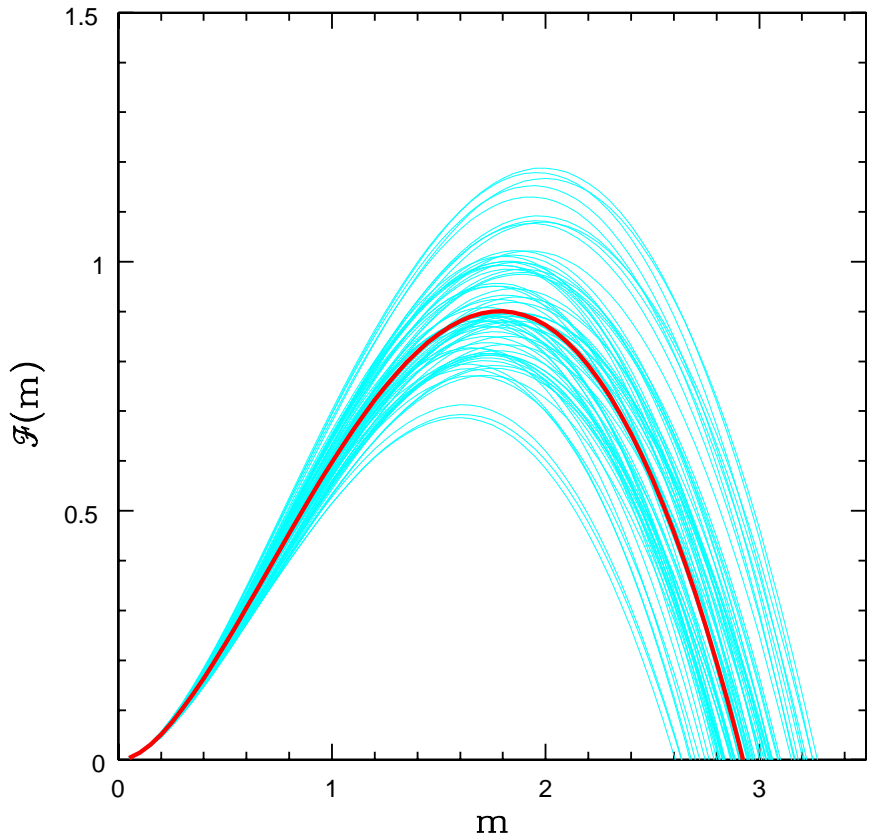}
\vskip-1.50truein
\caption{Relative mass fraction $\fom(m)$ as a function of stellar 
mass, where $m=M_\ast/M_\odot$. The thick red curve shows the
relative mass for typical values of the core formation time and disk
lifetime, which are functions of stellar mass as outlined in the
text. The collection of cyan curves shows the corresponding function
where the baseline parameters $\tcorz$ and $\tau_1$ are allowed to 
vary with a (uniform) random distribution and an amplitude of 10 
percent. }
\label{fig:massdep} 
\end{figure}

The relative mass fraction $\fom(m)$ from equation (\ref{massint})
is shown in Figure \ref{fig:massdep} for standard values of the input
parameters (solid red curve) and corresponding cases where the
parameters are allowed to vary (lighter cyan curves). All of the
curves follow the same basic trend: At low stellar masses, the
accretion rate through the disk ($\mdot\sim m^2$) becomes small, and
this trend leads to small relative mass scales and hence the partial
suppression of giant planet formation. In the opposite limit of large
stellar mass, $m\sim3$, the relative mass becomes small due to the
lack of time for accretion. The time required for core formation
becomes larger, while the disk lifetimes become shorter, so that
little time is left for the cores (planets) to accrete gas. The
observed scaling laws thus imply that larger stars ($m\gta3$) have
difficulty producing giant planets via the core accretion paradigm. In
the intermediate regime, for $m\sim0.5-2$, the relative mass
fraction varies (almost) linearly with increasing mass. This result
suggests that distribution of companion mass ratios should be nearly
the same for stars in this mass range. In constrast, the distribution
should be steeper for stellar masses outside this range ($m\ll0.5$
and/or $m\gg2$).

For completeness, we can also evaluate the effects of stellar mass
dependence in the limit where the disk mass accretion rate does not
vary appreciably during the epoch when the planet accretes most of 
its mass. This limit corresponds to large values of the time scale 
$t_0$. One can show that in the limit of large $t_0$, the quantities
from equation (\ref{massint}) become 
\be
M_0 \equiv \epsilon \mdotdz \tau_1 \qquad {\rm and} \qquad 
\fom(m) \equiv m \left[ 1 - {\tcorz\over\tau_1} m^{3/2} \right] \,. 
\label{massint2} 
\ee
In this case, the function $\fom(m)\to0$ for stellar mass
$m=(\tau_1/\tcorz)^{2/3}$, which is the same as found 
above.\footnote{This equivalence is expected. This point in parameter
space corresponds to the core formation time becoming larger than the
disk lifetime, and the crossover point is the same in both
approximations.}

\begin{figure}
\includegraphics[scale=0.70]{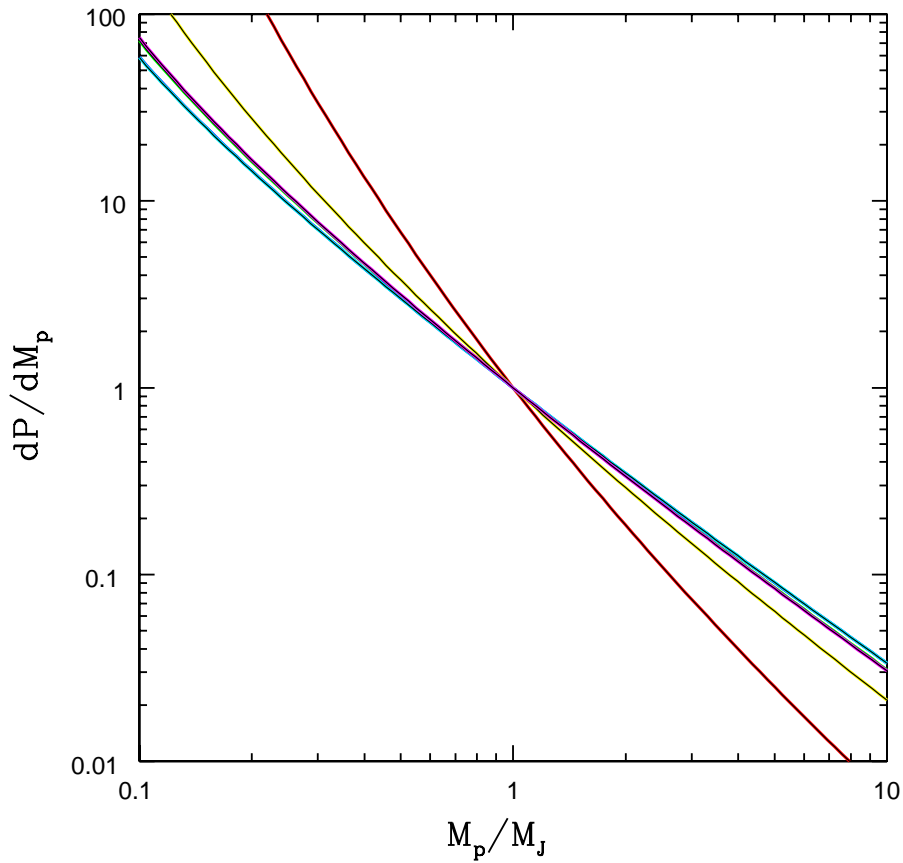}
\vskip-1.50truein
\caption{Planetary mass function for varying masses of the host 
star. The characteristic mass scale for the PMF is determined by the
available mass supply, which depends on disk masses, accretion rates,
and time scales according to equations (\ref{massint}) and 
(\ref{mzerodef}). Curves are shown for stellar masses of $m$ = 0.25
(red), 0.5 (yellow), 1.0 (green), 1.5 (blue), 2.0 (cyan), and 2.5
(magenta). All of the mass distributions are normalized so that they
are equal at $\mplan=\mjup$. }
\label{fig:star} 
\end{figure}

\begin{figure}
\includegraphics[scale=0.70]{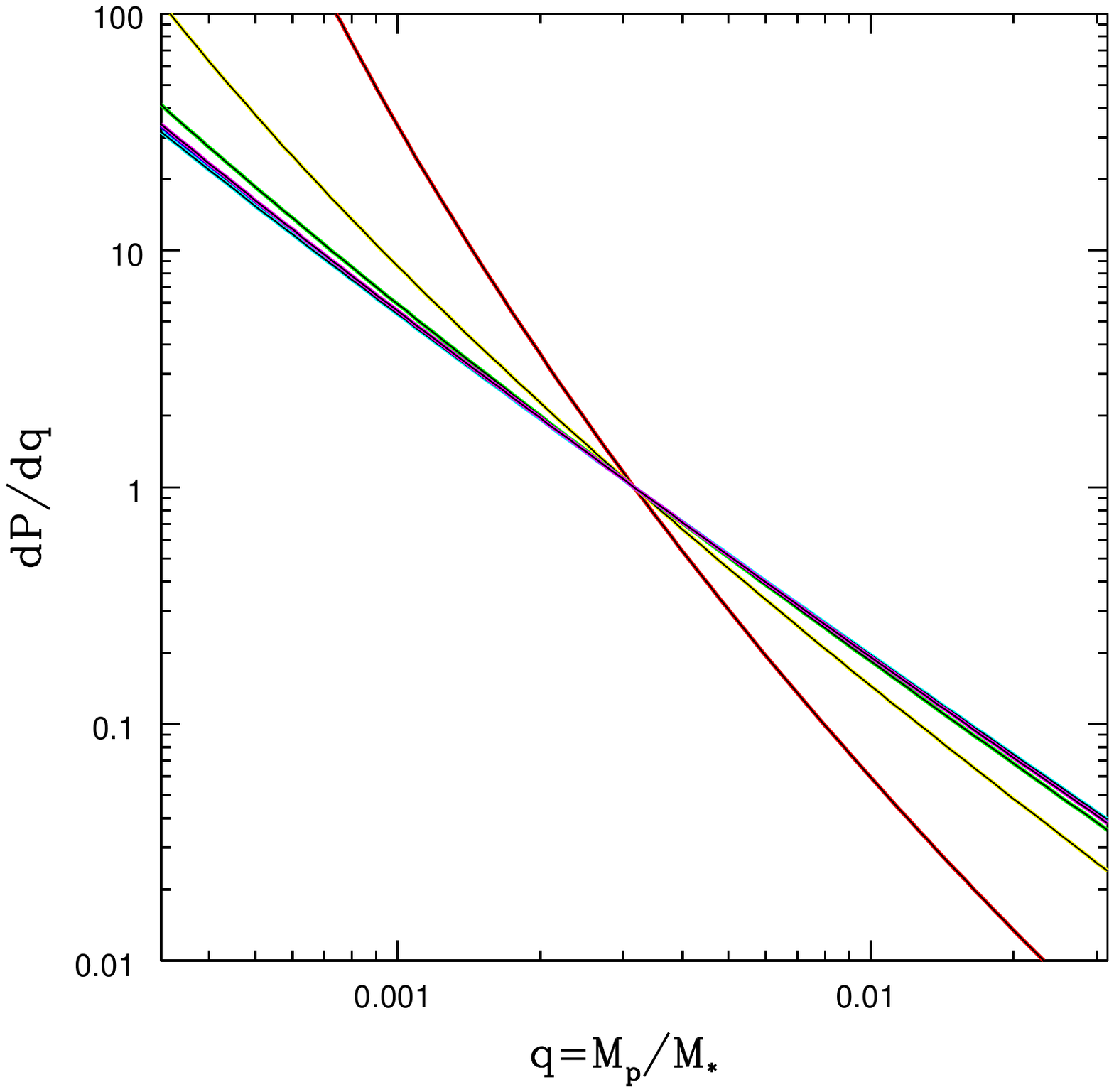}
\vskip-1.50truein
\caption{Planetary mass function expressed in terms of companion 
mass ratio $q=\mplan/M_\ast$ for varying masses of the host star. 
The mass scale $M_0$ for the PMF is determined by the available 
mass supply, which depends on disk masses, accretion rates, and 
time scales according to equations (\ref{massint}) and 
\ref{mzerodef}). Curves are shown for stellar masses of $m$ = 0.25
(red), 0.5 (yellow), 1.0 (green), 1.5 (blue), 2.0 (cyan), and 2.5
(magenta).}
\label{fig:qfun} 
\end{figure}

We can now construct the planetary mass function for different values
of the stellar mass. The stellar mass dependence is incorporated by
assuming that the total available mass for making a giant planet is
proportional to the relative mass fraction $\fom(m)$ defined
above. For given stellar mass, the characteristic mass scale $M_0$
that determines the shape of the PMF is then taken to have the form 
\be
M_0 = \fom(m) \times 5 \mjup \,. 
\label{mzerodef} 
\ee
Note that we are assuming that the mass scale for forming giant
planets is proportional to the total mass available. However, since
the disk can in principle form multiple planets, the coefficient in
equation (\ref{mzerodef}) is chosen to be only a fraction of the disk
mass. If the mass scale $M_0$ is used in the PMF constructed according 
to Section \ref{sec:models}, we obtain the distributions shown in
Figure \ref{fig:star}. The figure shows the mass functions for planets
forming around stars with masses $m$ = 0.25 -- 2.5. The planetary mass
distributions for host stars in the range $m$ = 1 -- 2.5 display a
similar shape. In contrast, stars with lower masses show a significant
deficit of larger planets, as shown by the PMF for $m=0.5$ (yellow
curve) and especially that for $m=0.25$ (red curve). These steeper
distributions imply that red dwarfs are less likely to have large
planets, but are still expected have some Jovian companions (see also
the more detailed planet formation models of \citealt{lba}).

For even larger host stars, the available mass supply decreases due to
the longer core formation times and shorter disk lifetimes (see Figure
\ref{fig:massdep}). Since $\fom(m)\to0$ for $m\gta3$, this present
framework does not provide a mass distribution for planets around
higher mass stars: such planets should be rare. This implication is
roughly consistent with current observational results \citep{reffert},
which find that the planet occurence rate drops significantly for
stellar masses larger than $M_\ast\sim2.5-3M_\odot$.

Emerging observational trends \citep{newmeyer} comparing low mass
brown dwarf companions to gas giant planets, as a function of stellar
mass, suggest that distributions of companion mass ratios are more
universal than the distributions of companion masses themselves.  In
addition to the PMF, we thus consider the distribution of the mass
ratios $q=\mplan/M_\ast$ in Figure \ref{fig:qfun} (for the same host
star masses as before; compare with Figure \ref{fig:star}). The
resulting distributions of mass ratio are roughly similar across the
range of stellar masses for $m\ge0.5$.  Once again, however, the
distribution for stars with the smallest mass, $m=0.25$ (red curve),
is steeper and shows a deficit of large-$q$ planets compared to the
others.  This deficit is due to the small fiducial mass scale, which
results from the smaller available mass supply in the disks
surrounding low mass stars. Although the curves are shown down to
ratios $q=3\times10^{-4}$ for all stellar masses, for $m=0.25$ this
value corresponds to a planet with only $\mplan\sim3$ $\mearth$. 
Planets with such small masses can accrete little gas, so that their
production mechanism lies outside this framework for the formation of
gas giants. Finally, we note that the similarity between the PMFs and
the mass-ratio distributions (Figures \ref{fig:star} and
\ref{fig:qfun}) arises because the distributions have a nearly
power-law form. In the limit where the PMF is exactly a power-law, the
two distributions would have the same shape.

This finding has observational implications. Microlensing surveys
measure companion mass ratios, and favor low mass stars (which are
more common than larger stars). On the other hand, both RV and transit
surveys have historically used our solar system as a template and
favored solar-type stars. As a result, the observed distribution of
companion mass ratios is predicted to be somewhat steeper for
microlensing surveys than for the others. For the largest stellar mass
considered here, $m=2.5$ (magenta curve), the PMF is nearly the same
as for solar type stars, but even higher mass stars ($m>3$) should
have relatively fewer large planets (because the relative mass
fraction $\fom\to0$).  Future observational work should determine how
the distributions of planetary masses and mass ratios vary across the
entire spectrum of stellar masses.

\section{Conclusion} 
\label{sec:conclude} 

This paper constructs a series of models for the initial mass function
for gas giant planets forming through the core accretion paradigm,
including its dependence on the mass of the host star. This section
provides a brief summary of our results (Section \ref{sec:summary}),
followed by a discussion of their implications (Section
\ref{sec:discuss}).

\subsection{Summary of Results} 
\label{sec:summary} 

This paper develops a working framework for determining the planetary
mass function within the core accretion paradigm. In this regime, the
final planet mass is determined by its mass accretion history (see
equation [\ref{mbasic}]). The treatment is semi-empirical in that the
distributions of the input parameters are specified (in part) using
observations of circumstellar disk properties.  This approach can
produce mass distributions that are roughly consistent with current
observations over the mass range $\mplan=0.1-10\mjup$ (e.g., see
Figure \ref{fig:pmfe1}).

Within current observational uncertainties, disk lifetimes are
observed to be exponentially distributed, whereas the planetary mass
function has a nearly power-law form. The framework developed in this
paper produces a nearly power-law PMF using the exponential
distribution of disk lifetimes as input. In order to obtain results
consistent with observations, other input parameters --- in addition
to the disk lifetime --- must sample distributions of values, and the
mass accretion rate must increase as the planet grows. In the limiting
case of a constant mass accretion rate, the exponential distribution
of disk lifetimes imprints an exponential fall-off in the PMF, which
leads to a deficit of larger planets compared to the observed
distribution.

This analysis reveals an interesting feature for the expected case
where the total disk lifetime has an exponential distribution: If we
consider the time remaining after some initial temporal offset, for
example after the planet reaches the stage where runaway gas accretion
occurs, then the distribution of time remaining (after the offset)
also has an exponential form (see equations [\ref{twigdef} --
  \ref{twigdist}]). Moreover, the exponential distribution has the
same decay constant or half-life.

In this formulation of the problem, flow of material onto the planet
is separated into two parts. The circumstellar disk provides a
background reservoir of gas that ultimately supplies the forming
planet.  In general, disks support a (slow) radially inward mass
accretion flow, and some fraction of this large-scale flux is
intercepted by the planet, i.e., enters its sphere of influence as
delineated by the Hill radius. In the vicinity of the planet, only a
fraction of the incoming material is actually accreted by the planet.
The overall efficiency of accretion onto the planet $\xi$, along with
the disk lifetime $t$, vary from case to case and provide sufficient
complexity to produce a nearly power-law PMF (Figure \ref{fig:pmfe1}).

Disk properties --- including lifetimes, mass accretion rates, and
expected/inferred core formation times --- are observed to vary with
the mass $M_\ast$ of the host star. These variations imply that the
total mass provided by the disk depends on stellar mass.  In this
approach, the fiducial mass scale $M_0$ for the PMF is assumed to
scale with the relative mass fraction $\fom$, thereby leading to
planetary mass distributions that depend on $M_\ast$ (Section
\ref{sec:varymass}).

The PMFs constructed in this paper indicate that the probability of
producing large planets around low mass stars is lower than for the
case of solar-type stars, i.e., the PMF is steeper for small stars
(Figure \ref{fig:star}). This finding is consistent with observational
surveys \citep{bonfils2013,vigan2020} as well as theoretical
expectations \citep{lba}). On the other hand, for sufficiently massive
stars, with $M_\ast\gta3M_\odot$, the core formation time scales
become comparable to typical disk lifetimes, and the probability of
producing giant planets is reduced (Figure \ref{fig:massdep}; see also
\citealt{reffert}). If we consider the distribution of mass ratios
$q=\mplan/M_\ast$, instead of the PMF itself, the resulting
distributions are roughly similar across the range of intermediate
stellar masses $M_\ast=0.5-2.5M_\odot$ (Figure \ref{fig:qfun}).
However, the mass ratio distribution for low mass stars remains
steeper than that for larger stars.

\subsection{Discussion} 
\label{sec:discuss} 

This paper constructs models for the PMF using a combination of disk
lifetimes and disk accretion properties. The microphysics of gas
accretion onto forming planets provides additional degrees of freedom.
Using the observed distribution of disk lifetimes and uniform
distributions for efficiency parameters that describe the accretion
processes, this framework can produce models that are roughly
consistent with the observed PMF. In addition, apparent trends
observed in the PMF for varying stellar mass can be reproduced.  Given
the current state of both observations and theory regarding the
distribution of planetary masses, however, this work represents only a
first step toward a complete understanding of the PMF. Many issues
should be addressed in greater detail.

The current version of the theory is incomplete. For example, this
paper accounts for the accretion history of gas flow onto forming
planets in a parametric manner, including the introduction of
efficiency factors. As the process of disk accretion becomes better
understood \citep{leebook,zhu2011,szulagyi}, the fundamentals of disk
physics should be incorporated into the models, ultimately leading to
specific probability distributions for the input variables. Another
complication is that several mechanisms can act to stop (or at least
slow down) accretion onto forming planets.  These processes include
gap opening in the circumstellar disks, the effects of the rotating
reference frame on accretion onto the planet surface, repulsion of the
accretion flow by magnetic fields, and the physics of accretion within
the circumplanetary disk. This current treatment also ignores
planetary migration, which can induce variations in the accretion rate
as the planets change their orbital elements (particularly if
accretion rates vary with radial location).  In addition, disks often
produce multi-planet systems, and the effects of additional bodies can
influence the formation of their neighbors. All of these effects
should be integrated into future studies.

The models considered here are relatively simple, in that they involve
only the disk lifetime $t$, an accretion efficiency factor $\xi$, and
the manner in which accretion scales with planetary mass.  As the
number of physical processes contributing to the determination of
planet mass increases, thereby including more variables that sample
distributions of values, the resulting PMF departs further from the
functional forms of the individual distributions. In the limit where
the number of independent contributing variables is large, the central
limit theorem comes into play, and the resulting PMF approaches a
log-normal form (e.g, \citealt{richtmyer}). Since the observed PMF is
approximately a power-law, this feature argues for an intermediate
number of relevant (and independent) variables. The process is not 
completely scale-free, however, and one should keep in mind that the 
power-law form of the PMF only manifests itself over a limited range 
of parameter space. 

The theoretical framework of this paper highlights two observational
issues that are important for understanding the PMF --- the
distribution of disk lifetimes for ages $t\sim10$ Myr and the
distribution of planetary masses for $\mplan\sim10\mjup$. The
characteristic mass scale $M_0$ is important for determining the shape
of the PMF (see equation [\ref{mzero}]), which depends on the time
scale $\tau$ appearing in the distribution of disk lifetimes.
Specification of the scale $\tau$ determines the fraction of
`long-lived' disks, which in turn determines the probability of
producing large planets with $\mplan\sim10\mjup$.  As a result, it
will be useful for observations to measure the fraction of disks that
live for $\sim10$ Myr \citep{than2018}.  If the value of the mass
scale $M_0$ is small (e.g., due to a small time scale $\tau$), the
formation of large planets is suppressed. The degree of suppression,
in turn, affects the apparent slope of the PMF for planets more
massive than Jupiter. The observational determination of the planetary
mass function for masses $\mplan\sim10\mjup$ is thus crucial for
testing the PMF produced by the core accretion paradigm (although such
observations will be challenging).

In addition to the core accretion paradigm, some planets could be
produced through gravitational instability in sufficiently massive
disks (see, e.g., \citealt{boss1997} to \citealt{wagner2019}). These
planets tend to have larger masses $\mplan\sim10\mjup$ (see, e.g.,
\citealt{abenz} to \citealt{inutsuka}) and could thus contribute to
the planetary mass function at the high-mass end. The results of this
paper suggest that the exponential distribution of disk lifetimes can
cause the core accretion paradigm to underproduce high mass planets,
so that additional planets produced via gravitational instability
could fill this gap. Notice also the brown dwarf binaries can
contribute to the `planetary' mass distribution for
$\mplan\sim10\mjup$ \citep{newmeyer}.

As outlined above, gravitational instability tends to produce more
massive planets, and they tend to form with large semi-major axes
(e.g., \citealt{rafikov}; see \citealt{vigan2017} for current
observational constraints). In contrast, the observed populations of
Hot Jupiters and more temperate gas giants tend to have smaller masses
then their colder counterparts (e.g., \citealt{howard2010}). As a
result, future studies (both observational and theoretical) should
determine the PMF for different ranges of orbital separations.

Finally, we note that theoretical descriptions of the planetary mass
function suffer from a lack of uniqueness. An intrinsic aspect of the
problem is that many physical processes contribute to the PMF, so that
the distributions of many input variables combine to produce a single
function. As a result, although the models of this paper successfully
reproduce the nearly power-law form of the PMF that is observed,
alternate approaches could also work. Fortunately, the framework
developed here is more robust than this initial application, so that 
future studies can improve the various steps of the calculation.

\bigskip 
$\,$ 
\noindent
{\bf Acknowledgment:} We would like to thank Veenu Seri for early work
that helped formulate the problem. We also thank Konstantin Batygin,
Nuria Calvet, and Greg Laughlin for many useful discussions, and thank
the referees for their comments on the manuscript.  This work was
supported by the University of Michigan, the NASA JWST NIRCam project
(contract number NAS5-02105), and the NASA Exoplanets Research Program
(grant number NNX16AB47G).

\appendix
\section{Planetary Mass Function for Constant Accretion Rate} 
\label{sec:appendix} 

In this Appendix, consider the limiting case where the mass 
accretion rate onto the planet approaches a constant and calculate 
the corresponding PMF. In this context, the 
planetary mass $\mplan$ is given by 
\be
\mplan = \mstart + \mdotz \xi t = \mstart + M \,,
\ee
where $t$ is the disk life time, $\xi$ is a uniform-random variable as
before, and the mass accretion rate $\mdotz$ is constant. The second
equality defines the mass $M$ accreted after the onset of rapid
accretion.

The distribution of the mass $M$ is determined by the 
cummulative probability 
\be
P(m<M) = \int_0^{t_\ast} {dt \over \tau} \exp[-t/\tau] + 
\int_0^{t_\ast} {dt \over \tau} \exp[-t/\tau] {t_\ast \over t} \,,
\ee
where the time scale $t_\ast$ is defined by 
\be
t_\ast = M/\mdotz \,. 
\ee
The probability distribution function for the accreted mass 
$M$ thus takes the form 
\be
{dP \over dM} = {1 \over \mdotz \tau} 
E_1 \left({M\over\mdotz\tau}\right) = 
{1 \over M_0} E_1 \left({M\over M_0}\right)\,,
\ee 
where the second equality defines $M_0=\mdotz\tau$. If the starting
mass $\mstart$ is constant, then the corresponding probability
distribution for the planetary mass $\mplan$ has the form
\be
{dP \over d\mplan} = {1 \over M_0} 
E_1 \left({\mplan-\mstart\over M_0}\right)\,. 
\ee

\end{document}